\documentclass[iop]{emulateapj}
\usepackage{graphicx}
\usepackage{multirow}
\usepackage{textcomp}
\usepackage{epsfig}
\usepackage{amsmath}
\usepackage{amssymb}
\usepackage{amsthm}
\usepackage{xy}

\slugcomment{Accepted 2016 August 12}

\def\xmm {\emph{XMM--Newton}}
\def\cxo {\emph{Chandra}}
\def\swift {\emph{Swift}}

\def\nustar {\emph{NuSTAR}}

\def\src {\mbox{1E\,161348--5055}}
\def\flux {\mbox{erg cm$^{-2}$ s$^{-1}$}}
\def\ergs {\mbox{erg s$^{-1}$}}
\def\nh {$N_{\rm H}$}
\def\cmdue{cm$^{-2}$}

\def\arcsec {$^{\prime\prime}$}
\def\arcmin {$^{\prime}$}

\def\3xmm{3XMM\,J1852$+$0033}

\shortauthors{N. Rea et al.}

\begin{document}

\title{Magnetar-like activity from the central compact object in the SNR RCW103}

\author{N. Rea\altaffilmark{1,2}, A. Borghese\altaffilmark{1}, P. Esposito\altaffilmark{1}, F. Coti Zelati\altaffilmark{1,3,4}, M. Bachetti\altaffilmark{5}, G. L. Israel\altaffilmark{6}, A. De Luca\altaffilmark{7}}
\altaffiltext{1}{Anton Pannekoek Institute for Astronomy, University of Amsterdam, Postbus 94249, NL--1090 GE Amsterdam, the Netherlands. Email: rea@ice.csic.es. The first four authors equally contributed to this work.}
\altaffiltext{2}{Institute of Space Sciences (IEEC--CSIC), Carrer de Can Magrans S/N, 08193 Barcelona, Spain.}
\altaffiltext{3}{Universit\`a dell'Insubria, via Valleggio 11, I--22100 Como, Italy}
\altaffiltext{4}{INAF--Osservatorio Astronomico di Brera, via Bianchi 46, I--23807 Merate (LC), Italy}
\altaffiltext{5}{INAF--Osservatorio Astronomico di Cagliari, via della Scienza 5, I--09047 Selargius (CA), Italy}
\altaffiltext{6}{INAF--Osservatorio Astronomico di Roma, via Frascati 33, 00040, Monteporzio Catone (RM), Italy}
\altaffiltext{7}{INAF--Istituto di Astrofisica Spaziale e Fisica Cosmica Milano, Via E. Bassini 15, I-20133 Milano, Italy}

\begin{abstract}

The 6.67\,hr periodicity and the variable X-ray flux of the central compact object (CCO) at the center of the SNR RCW\,103, named \src, have been always difficult to interpret within the standard scenarios of an isolated neutron star or a binary system. On 2016 June 22, the Burst Alert Telescope (BAT) onboard \swift\ detected a magnetar-like short X-ray burst from the direction of \src, also coincident with a large long-term X-ray outburst. Here we report on \cxo, \nustar, and \swift\, (BAT and XRT) observations of this peculiar source during its 2016 outburst peak. In particular, we study the properties of this magnetar-like burst, we discover a hard X-ray tail in the CCO spectrum during outburst, and we study its long-term outburst history (from 1999 to July 2016). We find the emission properties of \src\, consistent with it being a magnetar. However in this scenario, the 6.67\,hr periodicity can only be interpreted as the rotation period of this strongly magnetized neutron star, which therefore represents the slowest pulsar ever detected, by orders of magnitude. We briefly discuss the viable slow-down scenarios, favoring a picture involving a period of fall-back accretion after the supernova explosion, similarly to what is invoked (although in a different regime) to explain the ``anti-magnetar" scenario for other CCOs.

\end{abstract}


\section{Introduction}

The central compact object (CCO) \src, laying within the supernova remnant (SNR) RCW\,103, has been a mysterious source for several decades (Tuohy \& Garmire 1980; Gotthelf, Petre \& Hwang 1997). Despite presumably being an isolated neutron star (NS), it shows long-term X-ray outbursts lasting several years, where its luminosity increases by a few orders of magnitude. This source also has a very peculiar $\sim6.67$\,hr periodicity with an extremely variable profile along different luminosity levels (De Luca et al. 2006). Several interpretations of the nature of this system have been proposed, from an isolated slowly spinning magnetar with a substantial fossil-disk, to a young low mass X-ray binary system, or even a binary magnetar, but none of them is straightforward, nor can they explain the overall observational properties (Garmire et al. 2000; De Luca et al. 2006, 2008; Li 2007; Pizzolato et al. 2008; Bhadkamkar \& Ghosh 2009; Esposito et al. 2011; Liu et al. 2015; Popov, Kaurov \& Kaminker 2015).

A millisecond burst from a region overlapping the SNR RCW\,103 triggered the \swift\, Burst Alert Telescope (BAT) on 2016 June 22 at 02:03 UT (D'A\`i et al. 2016). These short X-ray bursts are distinguishing characteristics of the soft gamma repeater (SGR) and anomalous X-ray pulsar (AXP) classes, believed to be isolated NSs powered by the strength and instabilities of their $10^{14-15}$\,G magnetic fields (aka magnetars; Duncan \& Thompson 1992; Olausen \& Kaspi 2014; Turolla, Zane \& Watts 2015). In this work, we report on the analysis of the magnetar-like burst detected by \swift-BAT, on simultaneous \cxo\ and \nustar\ observations performed soon after the BAT burst trigger, and on the long-term \swift-XRT monitoring (\S\,\ref{obs}). Furthermore, we put our results in the context of all \swift, \cxo, and \xmm\ campaigns of \src\ from 1999 until 2016 July (\S\,\ref{results}). We then discuss our findings and derive constraints on the nature of this puzzling object (\S\,\ref{discussion}).

\section{X-ray observations and data analysis}
\label{obs}

\subsection{\swift}
\label{swift}

\begin{figure}
\centering
\includegraphics[width=5cm,height=10cm,angle=0]{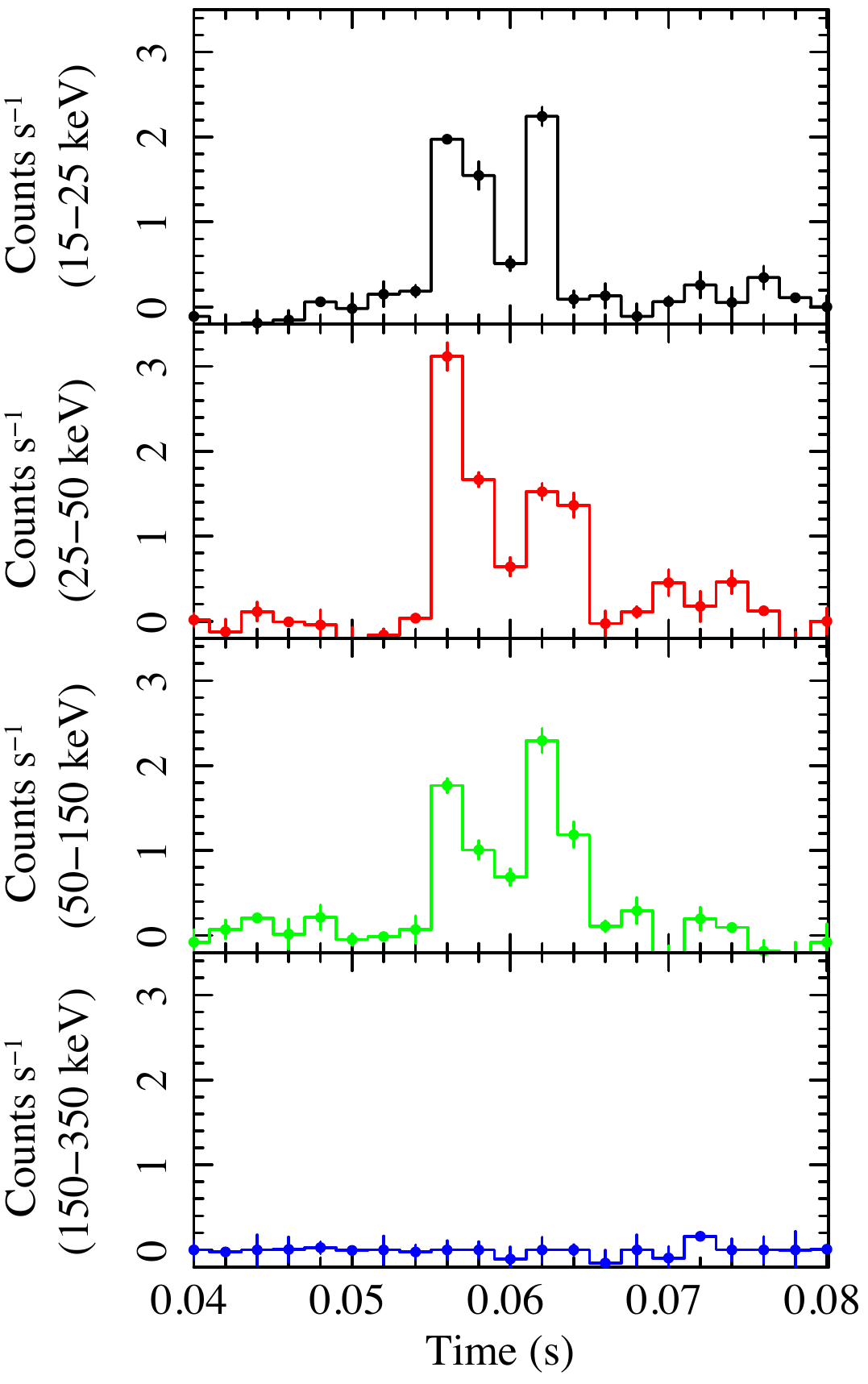}
\hspace{0.5cm}
\includegraphics[width=10cm,angle=0]{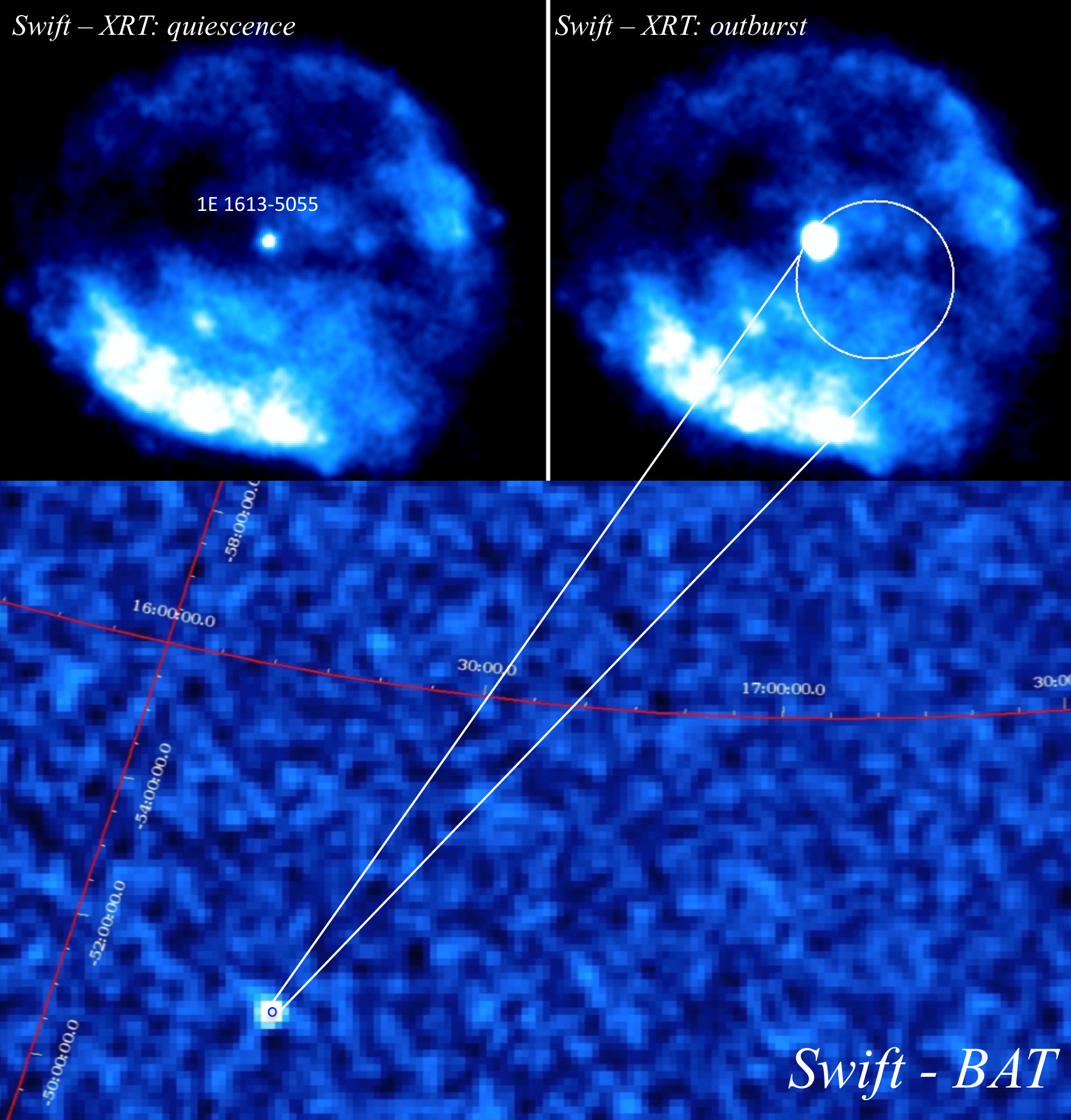}
\caption{Left panel:  \swift-BAT burst light curves at different energies (bin size: 2~ms). Right panel: \swift-BAT 15--150\,keV image of the burst detected on 2016 June 22 (bottom). Two \swift-XRT co-added 1-10\,keV images of the SNR RCW103 during the CCO quiescence state (from 2011-04-18 to 2016-05-16; exposure time $\sim66$\,ks; top-left) and outburst (from 2016-06-22 to 2016-07-20; exposure time $\sim67$\,ks; top-right). The white circle is the positional accuracy of the detected SGR-like burst, which has a radius of 1.5\arcmin\ (see text for details).}
\label{burst}
\end{figure}

The \swift\, X-ray Telescope (XRT) has been monitoring \src\, almost monthly, starting from April 2006 (Esposito et al. 2011). We have analyzed all \swift-XRT observations in photon counting mode from 2006 April 18 until 2016 July 20 (93 pre-burst and 20 post-burst observations, for an exposure of 236.2~ks). The last \swift-XRT observation prior to the burst was performed on 2016 June 22 from 01:30-01:42 UT (finished $\sim$20 min before the burst trigger), and showed the source already in an enhanced X-ray state (1--10 keV observed flux of $\sim 1.2 \times 10^{-10}$~\flux), while the previous observation was on 2016 May 16 from 13:47-15:47 UT with the source still at a low flux rate (1--10 keV observed flux of $\sim 1.7 \times 10^{-12}$~\flux; see Fig.\,\ref{burst} and \ref{longterm}).

The \swift\ data were processed and analyzed with usual procedures using the standard tasks included in the \textsc{heasoft} software package (v.6.19) and the calibration files in the 2016-05-02 \textsc{caldb} release. The \swift-XRT source counts were extracted from a circular  region centered on the most accurate position of the CCO ($\rm RA=16^h17^m36^s.23$, $\rm Dec= -51^\circ02'24''.6$; De Luca et al. 2008) with a radius of 10 pixels (1 pixel = 2.36\arcsec), and the background events from an annulus of radii of 10--20 pixels. Only the observation $\sim$20\,min before the BAT trigger, which yielded a severe pile-up, had to be extracted excising the inner 3.5 pixels of the extraction region.  

We analyzed the \swift-BAT data of the burst (trigger 700791, obs.ID 00700791000). The T90 duration of the event (the time during which 90\% of the burst counts were collected) was $0.009\pm0.001$~s and its total duration was $\sim$10~ms. These durations were computed by the Bayesian blocks algorithm \textsc{battblocks} on mask-weighted light curves binned at 1~ms in the 15--150~keV (Scargle 1998), where essentially all the emission is contained. For the burst only, mask-tagged light curves, images and spectra were created.  We extracted a 15--150-keV sky image and performed a blind source detection over the whole duration of the burst: a single, point-like source was detected at high significance (14.5$\sigma$) at the best-fit coordinates $\rm R.A. = 16^h17^m29\fs62,~Decl.=-51^\circ03'07\farcs9$, with an uncertainty radius of 1.5~arcmin (1$\sigma$, including a systematic error of 0.25~arcmin; Tueller et al. 2010). This position is consistent with a single known X-ray source: \src\, (see Fig.\,\ref{burst}).  No other X-ray source was detected within the burst error circle in the XRT data, with a $3\sigma$ 0.5--10\,keV detection limit of $<$0.003 counts~s$^{-1}$. Together with the exceptionally high flux of \src\ at the epoch of the burst, this strongly points to the CCO in RCW\,103 as the origin of the burst.

%%%%%%%%%%%%%%%%%%%%%%%%%%%%%%%%%
%%%%%%%%%%%%%%%%%%%%%%%%%%%%%%%%%%

\subsection{\cxo}
\label{chandra}

After the burst trigger, \src\, was observed with the Advanced CCD Imaging Spectrometer spectroscopic array (ACIS-S; Garmire et al. 2003) aboard the {\em Chandra X-Ray Observatory}, starting on 2016 June 25 at 09:20:07 until 22:00:38~UT, for an on-source exposure time of 44.2~ks (obs ID: 18878). The ACIS-S was configured in continuous clocking (CC) mode with FAINT telemetry format, yielding a readout time of 2.85~ms at the expense of one dimension of spatial information. The source was positioned on the back-illuminated S3 chip. 

We analyzed the data following the standard analysis threads\footnote{See \url{http://cxc.harvard.edu/ciao/threads/pointlike}.} with the \cxo\ Interactive Analysis of Observations software (\textsc{ciao}, v.~4.8; \textsc{caldb} v.~4.7.2). We accumulated the source photon counts within a box of dimension $3\times3$ arcsec$^2$ centered on the position of the CCO. The background was estimated by collecting photons within two rectangular regions oriented along the readout direction of the CCD, symmetrically placed with respect to the target and both lying within the remnant, whose spatial extension is $\sim9$ arcmin in diameter (Frank, Burrows \& Park 2015). The average source net count rate was $3.352 \pm 0.009$ counts~s$^{-1}$, which guarantees no pile up issues in the data set.

\begin{figure}
\centering
\resizebox{\hsize}{!}{\includegraphics[width=12cm,angle=0]{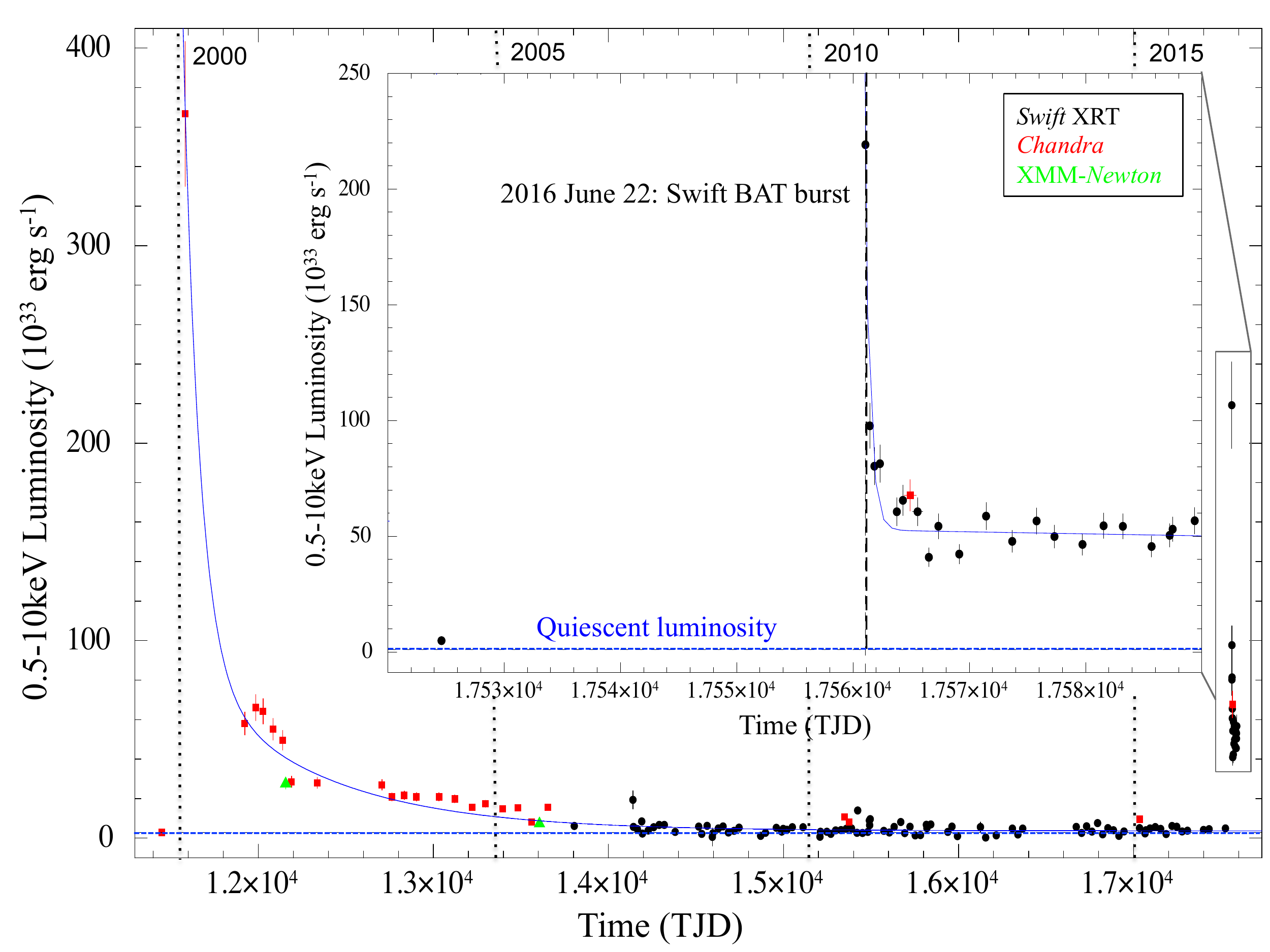}}
\caption{Long-term 0.5--10\,keV luminosity history of \src\, as observed since 1999-09-26 until 2016-07-20 by \cxo\,(red squares), \xmm\,(green triangles) and \swift\,(black circles). Dashed line represents the source quiescent luminosity. The inset is a zoom of the 2016 outburst.}
\label{longterm}
\end{figure}

We have also analyzed all archival \cxo\, observations pointing at $<$30\arcsec\, from our target (24 observations from 1999-09-26 until 2015-01-13; see Fig.\,\ref{longterm}). Photons from TE mode observations were extracted from a 2\arcsec\, circular region, and the background from an annulus with radii 4--10\arcsec. These observations were used for the timing and spectral long-term analysis (see below). When necessary, we corrected for pile up effects by using the model of Davis (2001).

\subsection{\nustar}
\label{nustar}

The {\em Nuclear Spectroscopic Telescope Array}  (\nustar; Harrison et al. 2013) observed \src\, starting on 2016 June 25 at 06:46:47~UT until June 26 at 18:42:50~UT, for a total on-source exposure time of 70.7\,ks (obs ID: 90201028002), simultaneously with the \cxo\, observation (\S\,\ref{chandra}). The data were processed using version 1.6.0 of the \nustar\, Data Analysis Software (\textsc{NuSTARDAS}) (using version 59 of the clock file to account for \nustar\ clock drifts caused by temperature variation). We used the tool \textsc{nupipeline} with default options for good time interval filtering to produce cleaned event files, and we removed time intervals corresponding to passages through the South Atlantic Anomaly. We ran the \textsc{nuproducts} script to extract light curves and spectra and generated instrumental response files separately for both focal plane modules (FPMA and FPMB). We collected the source counts within a circular region of 40\arcsec\, radius around the CCO position. The background subtracted source count rate in the 3-79\,keV was $0.27\pm0.03$ counts s$^{-1}$. We checked that a 30\arcsec\, extraction region gives consistent results. Background was estimated from two 60\arcsec\, circular regions in the same chip, one inside and one outside the ghost rays-contaminated area. We verified that the two background estimations did not significantly affect spectral modeling (see Fig.\,\ref{spectra}).

%%%%%%%%%%%%%%%%%%%%%%%%%%%%%%%%%
%%%%%%%%%%%%%%%%%%%%%%%%%%%%%%%%%%

\section{Results}
\label{results}

\subsection{Burst properties}

The light curve of the \swift-BAT burst shows a double-peak profile (Fig.\,\ref{burst}). We fit the spectra of the two peaks with single-component models typically used for magnetar bursts: a power law, a blackbody, and a bremsstrahlung component (e.g., Israel et al. 2008). Only the blackbody model provided an acceptable fit for both peaks. The first $\sim$5~ms of the event can be fit by a blackbody with $kT=9.2\pm0.9$~keV ($\chi^2_{\nu}$ = 1.03 (36 dof), null hypothesis probability (nhp) = 0.42), while for the second peak the blackbody temperature is $kT=6.0\pm0.6$~keV ($\chi^2_{\nu}$ = 1.22 (36 dof), nhp = 0.16). The corresponding total burst flux is $(1.6\pm0.2)\times10^{-6}$~\flux\ in the 15--150~keV range (corresponding to a luminosity of $2\times10^{39}$\ergs). All errors are given at 1$\sigma$ confidence level throughout the paper, and we assume a 3.3\,kpc distance (Caswell et al. 1975).

\subsection{Timing analysis}

\begin{figure*}
\centering
\hspace{-0.7cm}
\includegraphics[width=6.5cm,height=7.0cm,angle=0]{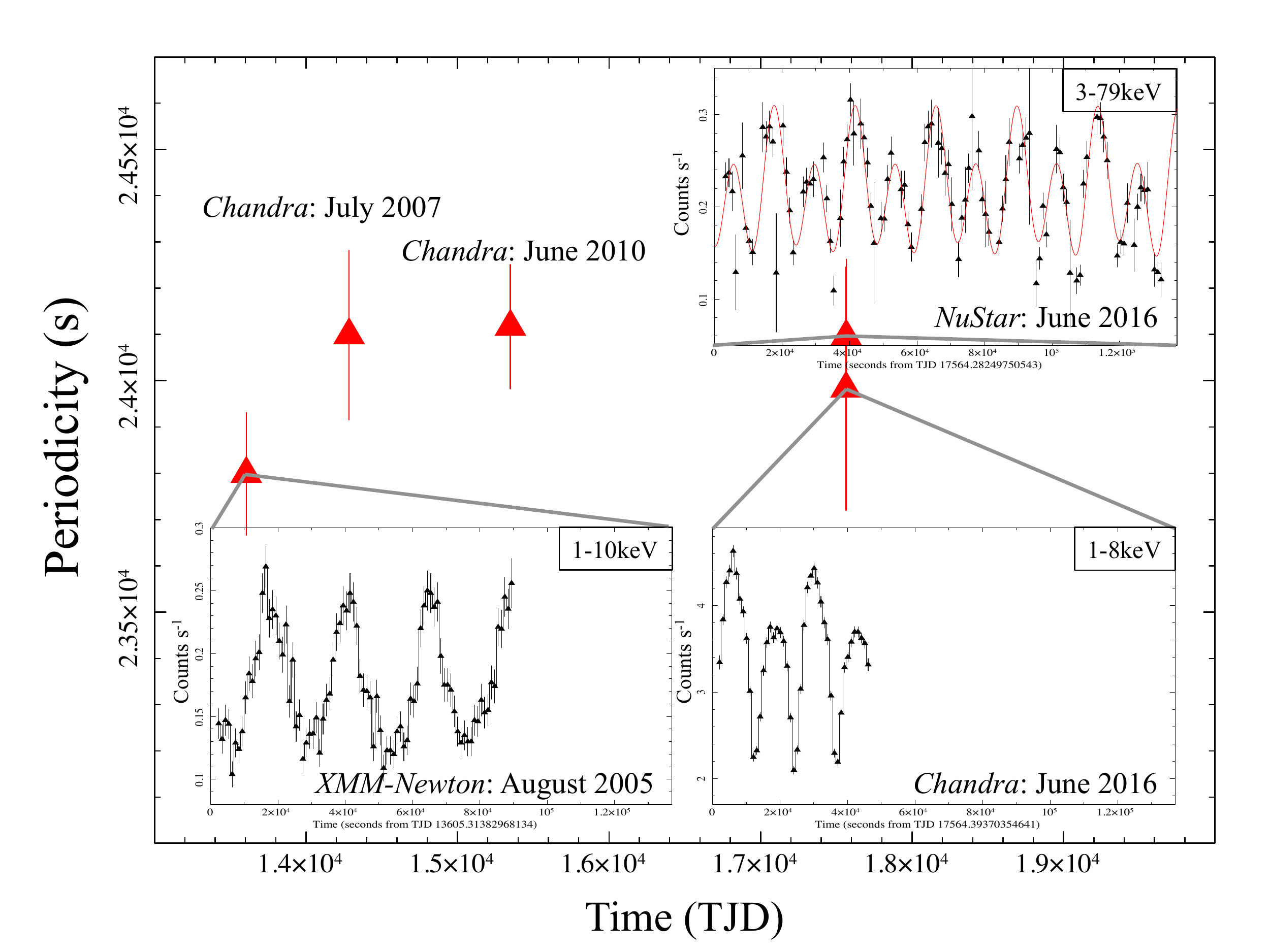}
\hspace{-0.4cm}
\includegraphics[width=5.5cm,height=6.7cm,angle=0]{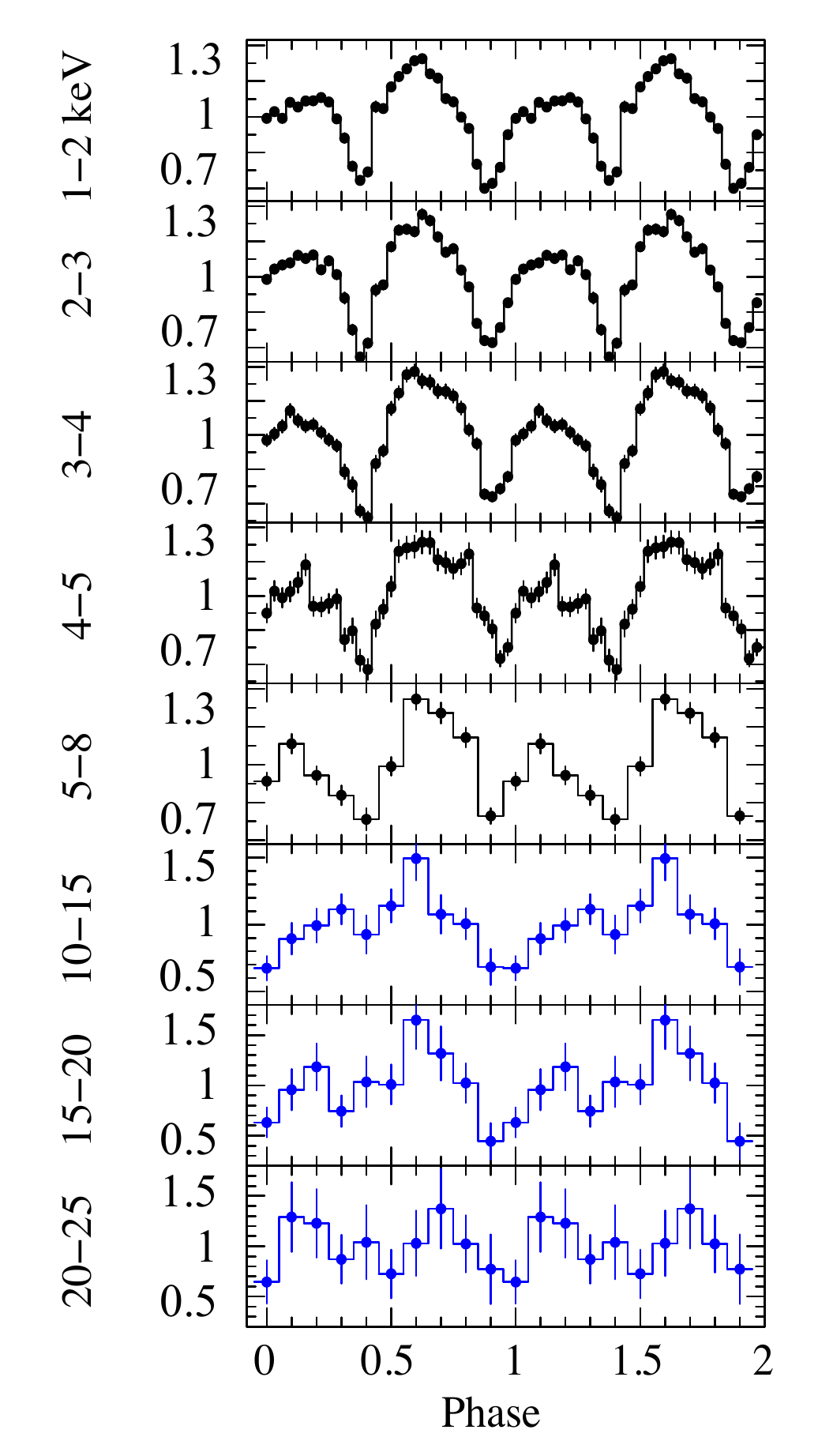}
\hspace{-0.3cm}
\includegraphics[width=6.8cm,height=6.8cm,angle=0]{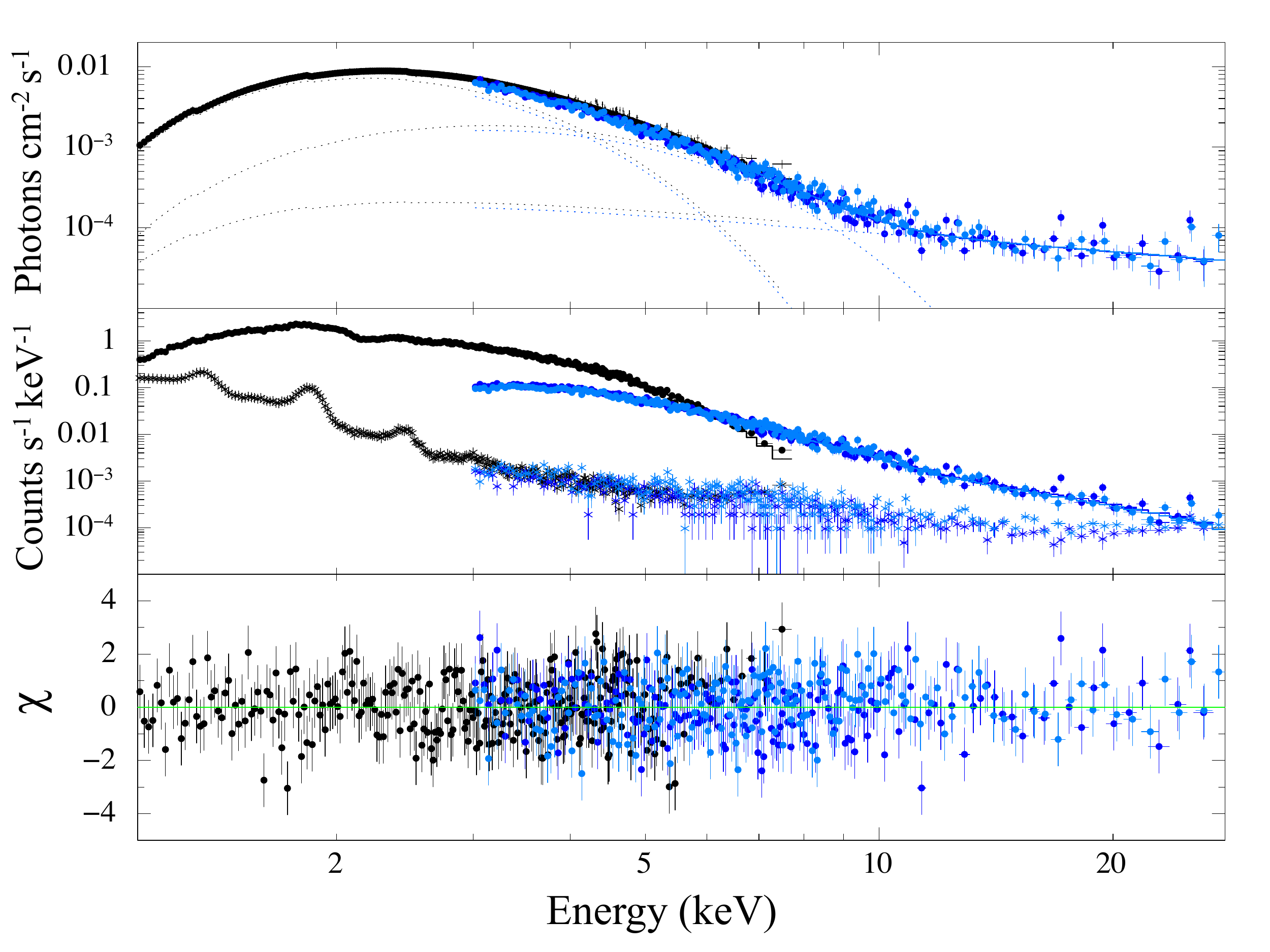}
\caption{{\em Left}: Period determination for the longest available archival X-ray observations, with the superimposed light curve binned at 1\,ks/bin. {\em Middle}: Energy-dependent, folded light-curve for the simultaneous \cxo\, (black) and \nustar\, (blue) observations soon after the 2016 burst. {\em Right}: Simultaneous spectral fit of the \cxo\,(black) and \nustar\,(light and dark blue) data with two absorbed blackbodies and a power-law component. The background spectrum is also plotted in the middle panel.}
\label{lc}
\label{spectra}
\end{figure*}

For the timing analysis, photon arrival times were reported to the Solar System barycentre frame, using the DE200 ephemerides and the \emph{Chandra} CCO position (see above).  We performed a blind search both for fast periodic and aperiodic signals using our new \cxo\, and \nustar\,data sets, using the \textsc{Xronos} timing package as well as the $Z_n^2$ test (Buccheri et al. 1983). We did not find any periodic signal via Fourier transform, but in both observations we detected the known $\sim$6.67\,hr periodic modulation (see Fig.\ref{lc}). We inferred 3$\sigma$ pulsed fraction upper limits (as explained in Israel \& Stella 1996), of 5\% (0.01--10 Hz), 6\% (10--100 Hz) and in the 7-9\% range for the highest sampled frequencies (100-200 Hz), for the \cxo\, observation. A similar analysis carried out on the \nustar\, data resulted on 3$\sigma$ upper limits of 12\% (0.01--3 Hz), and in the range 26--34\% at higher frequencies (3--1000 Hz).

In Fig.\,\ref{lc} we show the determination of the $\sim$6.67~hr period using the longest datasets in the X-ray archives, with the light curves of the two most extreme cases of a pure single peak (from \xmm\, in 2005; De Luca et al. 2006), and a clear double peak (in June 2016; this paper).  The 3-79\,keV light curve of the \nustar\, data and the simultaneous 1--8\,keV \cxo\, data were fit by two sinusoidal harmonics with fundamental periods 24095$\pm$167\,s (at TJD 17565.0) and 23983$\pm$263\,s (at TJD 17564.7), respectively.  We also studied the profile as a function of the energy in the 1--25\,keV band, and found that the profile may smooth to a single peak with increasing energy (see middle panel, Fig.\,\ref{lc}). 
Pulsed fractions (defined as the profile Max$-$Min/Max$+$Min) are: $40\pm1$\% in the 1--8\,keV \cxo\, band, and $41\pm5$\% for the 10--20keV \nustar\, data.

Studying the timing properties of \src\ is complicated by the changing pulse profile. However, if we assume the ephemeris of Esposito et al. (2011; solution `A'; constant period: 24\,030.42(2)~s; see the paper for a discussion of the assumptions and the validity of the solution), and extrapolate the phase of the minimum predicted for the fundamental harmonic, this is consistent within 2$\sigma$ ($\Delta\phi= 0.03\pm0.02$ for the 1--8~keV \cxo\, profile, and $\Delta\phi = 0.08\pm0.04$ for the 10--15\,keV \nustar\, profile) with that of the second minimum in Fig.\,\ref{lc}, around phase $\phi\sim0.9$ (implying $|\dot{P}|<7\times10^{-10}$~s s$^{-1}$).

%%%%%%%%%%%%%%%%%%%%%%%%%%%%%%%%%
%%%%%%%%%%%%%%%%%%%%%%%%%%%%%%%%%%

\subsection{Spectral analysis}

We started the spectral analysis (always using \textsc{xspec} v.12.9), by simultaneously fitting the new \cxo\, and \nustar\, observations (see Fig.\,\ref{spectra}). We found that although the \cxo\, spectrum alone is well fit with two blackbodies, this is not the case when taking into account also the \nustar\, hard X-ray spectrum of \src. A good fit is found for a model comprising of two absorbed (\nh$=2.05(5)\times10^{22}$~\cmdue) blackbodies with temperatures of kT$_1=0.52\pm0.01$\,keV and kT$_2=0.93\pm0.05$\,keV, with radii of R$_1=2.7\pm0.7$\,km and R$_2=0.4\pm0.2$\,km, plus the addition of a power-law component with photon index $\Gamma=1.20\pm0.25$ (adding a constant between the two instruments to account for inter-calibration uncertainties, which was always within 10\%). The total observed flux in the 0.5--30\,keV energy range is $(3.7\pm0.1)\times10^{-11}$~\flux, and the joint fit gives $\chi^2_{\nu}=1.04$ (660 dof; nhp = 0.2). A model with a blackbody plus two power-laws results in a slightly worse fit of $\chi^2_{\nu}=1.2$ (660 dof; nhp = $2\times10^{-4}$), and bad residual shape.

\subsection{Outburst history}

To study the outburst history of \src, we reanalyzed all the available \cxo, \xmm, and \swift\ data of the source acquired from 1999 until 2016 July (see Fig.\,\ref{longterm}). All spectra were fit by fixing the absorption column density to the value derived using \cxo\, (\nh$=2\times10^{22}$\cmdue) plus two blackbody components (because in the 1--8\,keV range the hard X-ray power-law is not required by the fit and contributes less than 10\% to the flux in this band). We show the extrapolated 0.5--10\,keV luminosity in Fig.\,\ref{longterm}. This source underwent two outbursts in the past $\sim$17\,years. The first outburst can be empirically fit by a constant plus three exponential functions, resulting in a total (impulsive plus persistent) emitted energy in the 0.5-10\,keV band of $E_{1st-out}\sim9.9\times10^{42}$\,erg. This outburst was characterized by heating of two different regions on the surface, with the two blackbodies in the X-ray spectra cooling and shrinking from the outburst peak until quiescence: $kT_1\sim$0.6--0.4\,keV ($R_1\sim5-1$\,km), and $kT_2\sim$1.4--0.7\,keV ($R_2\sim1.4-0.1$\,km). This new second outburst, that started $<$1~month before the SGR-like burst (see \S\ref{swift}), shows similar energetic and spectral decomposition so far ($E_{2nd-out}\sim1.6\times10^{42}$\,erg). Furthermore, our \nustar\, observation shows for the first time, that during the outburst peak this source emits up to $\sim30$\,keV (certainly modulated until $\sim20$\,keV; Fig.\,\ref{lc}).

\section{Discussion}
\label{discussion}

We report on the analysis of a magnetar-like short burst from the CCO \src\, (D'Ai et al. 2016), and study its coincident X-ray outburst activity. This short ms-burst and its spectrum, the X-ray outburst energetics of this source, the spectral decomposition, and surface cooling (see \S\,\ref{results}) are all consistent with observations of magnetar SGR-like bursts and outbursts (see Rea \& Esposito 2011, and reference therein, for an observational review). This is the second X-ray outburst detected from \src, and it shows for the first time a coincident SGR-like burst and a non-thermal component up to $\sim30$\,keV. Two-peak SGR-bursts with similar luminosity and spectra have been observed in other magnetars (see e.g. Aptekar et al. 2001, G\"otz et al. 2004, Collazzi et al. 2015). Due to their ms-timescales and relatively soft spectra, these events cannot be interpreted as Type\,I X-ray bursts or short GRBs (see Galloway et al. 2008; Sakamoto et al. 2011). On the other hand, hard X-ray emission has been detected for at least half of the magnetar population (Olausen \& Kaspi 2014). Sometimes this emission is steady, but other times transient and connected with the outburst peaks. Magnetar outbursts are expected to be produced by the instability of strong magnetic bundles which stress the crust (from outside or inside: Beloborodov 2009, Li, Levin \& Beloborodov 2016, Perna \& Pons 2011, Pons \& Rea 2012). This process heats the surface in one or more regions, and at variable depth inside the NS crust, which in turn drives the outburst duration. The high electron density in these bundles might also cause resonant cyclotron scattering of the seed thermal photons, creating non-thermal high-energy components in the spectrum. Such components can be transient if the untwisting of these bundles during the outburst decay produces a decrease in the scattering optical depth. Furthermore, magnetospheric re-arrangements are expected during these episodes, and are believed to be the cause of the short SGR-like bursts (see Turolla, Zane \& Watts 2015 for a review). Repeated outbursts on several-year timescales have also been detected in at least four magnetars (Bernardini et al. 2011; Kuiper et al. 2012; Archibald et al. 2015), and their recurrence time is expected to be related to the source magnetic field strength and configuration, and to the NS age (see Perna \& Pons 2011; Vigan\`o et al. 2013).

In this scenario, the only puzzling property of \src, that makes it unique among any SGR, AXP, CCO or other known NS, is the 6.67\,hr long periodicity, which would represent the longest spin period ever detected in a pulsar. On the other hand, the extreme variability of the modulation in time and energy strongly disfavor this modulation being due to an orbital period (see detailed discussion in De Luca et al. 2008, Pizzolato et al. 2008), but remain fully consistent with the usual pulse profile variability observed in actively flaring magnetars (see e.g. Rea et al. 2009, 2013; Rodr{\'{\i}}guez Castillo et~al. 2014).

Isolated pulsar spin periods are observed to be limited to $\sim$12\,s, with the slowest pulsars indeed being the magnetars. This period distribution is explained as due to Hall-Ohmic magnetic field decay during the evolution of these neutron stars (see Pons, Vigan\`o \& Rea 2013). The slowest isolated pulsar that magnetic field decay might produce is $\sim30-50$\,s, according to self-consistent 2D simulations (e.g. Vigan\`o et al. 2013), if we consider the generous case of field threading the stellar core, zero dissipation from crustal impurities, and an initial field ranging from 10$^{13-15}$\,Gauss, while using typical spin period at birth in the range of 1--300\,ms. Regardless of the model inputs, we can in no case reproduce hours-long spin periods.

Given the strong evidence for the magnetar nature of the X-ray emission of this source, we are now left with discussing all possible slow-down mechanisms other than the typical pulsar dipolar loss. Since its discovery, many authors have already discussed several scenarios (see De Luca et al. 2006; Li 2007; Pizzolato et al. 2008; Bhadkamakar \& Gosh 2009; Lui et al. 2015; Popov, Kaurov, \& Kamiker 2015), which we cannot summarize here. We will however highlight and discuss the possibilities that remain open, along with their possible deficiencies. 

The first possibility could be a long-lived fossil disk (Chatterjee, Hernquist \& Narayan 2000), which forms via the circularization of fall-back material after the supernova explosion (see i.e. De Luca et al. 2006; Li 2007). This might result in substantially slowing the spin period. However, recent studies on the formation of fossil disks apparently disfavor their existence around NSs under reasonable assumption on the magnetic torque in the pre-SN phase (Perna et al. 2014). On the other hand, the magnetar flaring activity during its lifetime would most probably expel such thin disks very quickly. 

Another possibility is that \src\, is a magnetar in a low mass X-ray binary with an M6 companion (or later; De Luca et al. 2008), emitting as though it were isolated, but that had its spin period tidally locked to the orbital motion of the system (see i.e. Pizzolato et al. 2008). However also in this case, fine-tuning is needed to explain how a very low-mass companion remains gravitationally bound to the magnetar after the SN explosion.

The most viable interpretation, in line with what has been proposed for other CCO systems (the ``anti-magnetars": see e.g. Halpern \& Gotthelf 2010; Torres-Forn\'e et al. 2016), seems to be of a magnetar that had a strong SN fall-back accretion episode in the past (Chevalier 1999). In particular, if \src\, is born with a magnetic field and spin period such that when the fall-back accretion begins, the source is in the propeller regime (Illarionov \& Sunyaev 1975; Li 2007; Esposito et al. 2011), then the accreted material will not reach the surface and bury the B-field, as for the "anti-magnetar" CCOs, but in the first years or more of its lifetime the magnetar will accrete onto the magnetosphere, hence with a substantially larger spin-down torque. When the fall-back accretion stops, the magnetar continues to evolve as any other isolated pulsar, but with a substantially slower spin period.

\section*{Acknowledgements} 

We are grateful to Dr. Belinda Wilkes and Dr. Fiona Harrison for accepting our Director Discretionary Time requests, and to the \cxo\, and \nustar\, teams for the large efforts in scheduling these simultaneous observations on such a short timescale. We also acknowledge the \swift\, team for promptly announce new transient events, allowing rapid follow-up observations. NR, AB, PE and FCZ acknowledge funding in the framework of the NWO Vidi award A.2320.0076 (PI: N.~Rea), and via the European COST Action MP1304 (NewCOMPSTAR). NR is also supported by grants AYA2015-71042-P and SGR2014-1073.  We acknowledge J. Pons and J. Elfritz for useful discussions and valuable comments, and the referee for his/her suggestions.

\end{document}